\documentclass[preprintnumbers,aps,prb,twocolumn]{revtex4}

\usepackage{graphicx}
\usepackage{dcolumn} 
\usepackage{amsmath}
\usepackage{exscale}
\usepackage{bm}      
\usepackage{color}
\usepackage{epsfig,psfrag,subfigure,subeqnarray}
\usepackage{bbm}
\usepackage{amsbsy}
\usepackage{amssymb}
\usepackage{enumitem}

\def\lan{\langle}
\def\ran{\rangle}

\def\vk{{\bf k}}
\def\vK{{\bf K}}

\def\vr{{\bf r}}

\def\vQ{{\bf Q}}

\def\vp{{\bf p}}

\newcommand{\bd}{\begin{equation}}
\newcommand{\ed}{\end{equation}}
\newcommand{\be}{\begin{equation}}
\newcommand{\ee}{\end{equation}}
\newcommand{\bt}{\begin{split}}
\newcommand{\et}{\end{split}}

\newcommand{\bn}{\begin{align}}
\newcommand{\en}{\end{align}}

\newcommand{\bea}{\begin{eqnarray}}
\newcommand{\eea}{\end{eqnarray}}
\newcommand{\ba}{\begin{array}}
\newcommand{\ea}{\end{array}}

\begin{document}

\title{Fundamental differences between exciton and quantum dot duo}

\author{Monique Combescot}
\affiliation{Sorbonne Universit\'e, CNRS, Institut des NanoSciences de Paris, 75005-Paris, France}
\author{Shiue-Yuan Shiau}
\affiliation{Physics Division, National Center for Theoretical Sciences, Hsinchu, 30013, Taiwan}
\author{Valia Voliotis}
\affiliation{Sorbonne Universit\'e, CNRS, Institut des NanoSciences de Paris, 75005-Paris, France}

\begin{abstract}
We present five major reasons why semiconductor exciton, that is, a correlated electron-hole pair in a bulk, quantum well, or quantum wire, is conceptually different from a pair in a quantum dot: (1) the origin of pair binding, (2) the interaction with additional carriers, (3) the quantum nature of the pair, (4) the coupling to photon, and (5) the photon-absorption mechanism. Due to these  differences, we should refrain from calling an electron-hole pair in a quantum dot  an exciton, as commonly done; we propose to call it a duo. Within the same frame of chamber musics, we likewise propose to call three and four carriers in a dot, a trio and a quatuor, instead of a trion and a biexciton.
\end{abstract}
\date{\today}

\maketitle

The impressive progress of growth techniques over the past decades has allowed making very high-quality low-dimensional semiconductor structures --- quantum wells, quantum wires and quantum dots --- opening new avenues of research and innumerable applications in optoelectronics and spintronics ready for quantum information technologies\cite{Bastard,spin,Michler}. As a direct consequence of quantum confinement, the elementary semiconductor excitations can reach very large binding energies, making  the field of so-called ``excitonics" a promising research area\cite{2D}. While the concept of exciton is meaningful when  spatial confinement is along one (quantum wells) or two (quantum wires) dimensions, we here show that the picture of a bound electron-hole pair as an exciton breaks down when the three spatial dimensions are confined (quantum dots). This is why, instead of calling an electron-hole pair in a quantum dot (QD)  an exciton, as we do for other structures, we should use another terminology. This problem is more than just about semantics; the physical understanding is totally different as to the pair  interacting with additional carriers and coupling to photons, and the possibility of photon absorption.

 Elementary excitation in undoped semiconductors consists in removing an electron from the valence band and bringing it to the empty conduction band, the energy associated with this excitation being of the order of the band gap. The full valence band with an empty state behaves for most physical effects as a single particle with a positive mass and a positive charge, that we call hole. Repeated Coulomb attraction between this valence hole and the conduction electron leads to a correlated pair state called exciton, with bound and unbound levels, very much like a hydrogen atom\cite{exciton}. As Coulomb interaction conserves momentum, the  electron-hole pairs that enter this repeated attraction have a constant momentum $\vk_e+\vk_h=\vQ$, which corresponds to the exciton center-of-mass  momentum. The exciton also has a  quantum index $\nu$ that differentiates its bound and unbound levels. So, the exciton creation operator $B^\dag_{\vQ,\nu}$ ends up reading in terms of electron and hole creation operators $(a^\dag_{\vk_e},b^\dag_{\vk_h})$ as 
\be
B^\dag_{\vQ,\nu}= \sum_{\vk_e,\vk_h} a^\dag_{\vk_e}b^\dag_{\vk_h}\lan \vk_e,\vk_h|\vQ,\nu\ran\,.\label{1}
\ee

\noindent The pair prefactor differs from zero for $\vk_e+\vk_h=\vQ$. To  grasp the consequences of having the exciton in a bound state, we can introduce the pair relative-motion momentum $\vp$, which for electron and hole masses $(m_e,m_h)$, is such that $\vk_e=\vp+\gamma_e \vQ$ and $\vk_h=-\vp+\gamma_h \vQ$ with $\gamma_e=1-\gamma_h=m_e/(m_e+m_h)$. This reduces the exciton creation operator to
\be
B^\dag_{\vQ,\nu}= \sum_{\vp} B^\dag_{\vQ,\vp}\lan \vp|\nu\ran\,\label{2}
\ee
with $B^\dag_{\vQ,\vp}=a^\dag_{\vp+\gamma_e \vQ}b^\dag_{-\vp+\gamma_h \vQ}$. Bound exciton states have a relative-motion wave function $\lan \vr|\nu\ran$ very localized in real space, or equivalently very extended in momentum space. The opposite holds for extended states. The spatial extension $a_X$ of a bound-state exciton is determined by the competition between the electron-hole Coulomb attraction $e^2/\epsilon_{sc}a_X$ where $\epsilon_{sc}$ is the semiconductor dielectric constant, and the kinetic energy cost $\hbar^2/2\mu_X a_X^2$  to localize the electron and hole within a distance $a_X$ from each other, $\mu_X^{-1}=m_e^{-1} + m_h^{-1}$ being the electron-hole relative-motion mass. Balance between these two energies leads to
\be
a_X\propto \frac{\hbar^2 \epsilon_{sc}}{e^2\mu_X }\,,\label{3}
\ee 
which is the physical scale for the distance at which the electron stays  to the hole as a result of this competition, that is, when there is at least one spatial dimension along which the carriers can move arbitrarily far away from each other.

The situation is essentially  the same when one spatial dimension is  confined, as for quantum wells. The component of the carrier momentum along the confined direction is no more quantized in $2\pi/L$ as for a size-$L$ bulk sample, but has a far larger quantization scale, in $2\pi/\ell$, where $\ell$ is the quantum well width. The major effect of a $\ell$ value small compared to $a_X$ is to  split the bulk exciton ground state. It also increases the exciton binding energy due to the reduction in charge screening, up to four times the bulk value\cite{Bastard}, with a Bohr radius consequently twice smaller, as mathematically obtained for two-dimensional systems, that is, for $\ell=0$. 

 The situation is more delicate when two spatial dimensions are confined, as for quantum wires, because if we shrink these two dimensions  to zero, the Schr\"{o}dinger equation becomes singular, with a binding energy that goes to infinity. So, the finite value of the pair binding energy strongly depends on the wire cross section. Here also, the spatial extension of the ground-state exciton results from a competition between Coulomb attraction, that keeps the electron close to the hole, and localization energy, but the wire confinement plays a significant role in the resulting extension\cite{Guillet}.

The effect of confinement is even more dramatic when the three spatial dimensions  are confined, as for quantum dots. Carriers are forced to be in a dot by the potential barriers, regardless of their Coulomb energy, even when there is a strong repulsion between same-charge carriers. So, in a QD, the pair extension is  controlled by the dot confinement\cite{Zunger, MAD}.
This has been substantially discussed in the case of semiconductor microcrystals\cite{Efros,Hanamura,Takagahara,Que}, or interface islands as a result of the thickness fluctuations of one monolayer in a quantum well\cite{Andreani}, and more recently in III-V self-assembled QDs\cite{Stobbe}. 

In the following, we will restrict to QDs in the strong confinement regime. In this regime, the very first fundamental difference between an electron-hole pair in a QD and a pair in a crystal  is the binding mechanism: confinement for a dot and Coulomb attraction for all the other cases that have at least one spatial dimension along which electron and hole can move far away. Just for this reason, an electron-hole pair in a dot should not be called an exciton. We propose to call it a dot duo, or just a \textit{duo}.

Of course, these different binding mechanisms go along with other fundamental differences that altogether constitute even stronger supports for us to refrain from calling an electron-hole pair in all sample geometries  an exciton. Let us discuss four other important differences: (i) the interaction with additional carriers, (ii) the quantum nature of the pair, (iii) the coupling to photon, (iv) the photon-absorption mechanism.

\textbf{(i) Interaction with additional carriers} 

The electric dipole induced by an electron-hole pair bound into an exciton can attract a free charge, that is, an electron or a hole, or the dipole of another exciton. Yet, in this attraction, the quantum nature of the particles enters into play through the fact that a bound state can result from this attraction, provided that the  spins of the same carrier species are different---which fundamentally originates from the Pauli exclusion principle.

As electrostatic forces come from Coulomb interaction, the spatial extension of the resulting multi-particle bound state also scales as $a_X$, but with a prefactor smaller than 1 because the exciton is globally charge neutral. Moreover, for these bound states to be determined through the competition between Coulomb attraction and localization energy, one spatial dimension at least has to be free of confinement, as for bulks, quantum wells, and quantum wires. These geometries have positively or negatively charged trions, $(ehh, eeh)$, or neutral biexcitons $(eehh)$; their  energies are lower than the energy of a free exciton plus a free carrier or a free exciton---otherwise these carriers would not stay close to each other\cite{Shiau}. This has been experimentally shown in the case of quantum wells\cite{trion,biexciton}.

For QDs, the principal force that brings the carriers together is not the Coulomb attraction but the barrier potential. As a result, the carriers can be constraint to stay at a distance small compared to the trion or biexciton spatial extension; in this case, there is no fundamental reason for two duos in a dot to have an energy lower than twice the energy of one duo in a dot. And actually, ``unbound biexcitons'' have been experimentally observed\cite{BX}. It also is interesting to note that an electron can lower its energy by being in a dot already hosting a duo, whereas a hole added to such a dot would increase its energy\cite{X-X+}. This sign change in ``binding energy'' physically comes from the fact that the electron and hole wave functions  leak out of the dot differently: while infinite barriers force the carriers to be inside the dot whatever their mass, the carriers have a non-zero probability to be outside the dot when the barriers are finite, and this probability depends on the carrier mass. In usual III-V heterostructures, the barrier height for electron is considered twice as large as the one for hole. Still, due to its much smaller mass, the electron leaks out of the dot more than the hole (see appendix). This makes the (overall neutral) duo positively charged inside the dot and negatively charged outside, which explains why an additional electron gains energy by being in a dot having a duo\cite{Finley}, while this costs energy to a hole\cite{Regelman}. Although the spectroscopic signature of differently charged complexes is very sensitive to the quantum dot shape, composition, and induced strain, this general trend stays valid for standard self-assembled InAs/GaAs quantum dots\cite{X-X+,Finley,Regelman}.

So, just as an electron-hole pair in a dot should not be called an exciton but a duo, we suggest to call three or four carriers in a dot, not a trion or a biexciton, but a \textit{trio} or a \textit{quatuor}.

\textbf{(ii) Quantum nature of the electron-hole pair}

Being made of fermion pairs, excitons are bosonic particles\cite{Combescotbook}. More precisely, they are composite bosons, as seen from their commutation relations, which for exciton eigenstate $i=( \vQ_i,\nu_i)$, read as
\bea
\left[B_m,B^\dag_i\right]_-&=& \delta_{mi}-D_{mi}\,,\label{4}\\
\left[B^\dag_i,B^\dag_j\right]_-&=&0\,.\label{5}
\eea

The fermionic components of the exciton appear through the $D_{mi}$ operator. This operator gives zero when acting on vacuum, while acting on other excitons, it generates  fermion exchanges that the excitons have between them in the absence of Coulomb processes, through the so-called Pauli scatterings $\lambda\left(\begin{smallmatrix}
n&j \\ m& i\end{smallmatrix}\right)$ defined as 
\be
\left[D_{mi},B^\dag_j\right]_-=\sum_n\Big(\lambda\left(\begin{smallmatrix}
n&j \\ m& i\end{smallmatrix}\right)+(i\longleftrightarrow j)\Big)B^\dag_n\,. \label{6}
\ee

The Pauli exclusion principle between the exciton components also appears in Eq.~(\ref{5}), but in a more subtle way. From this equation, we readily get $(B_0^\dag)^2\neq0$, while for fermions, we  would have $(a_{\bf0}^\dag)^2=0$. This is why $N$ excitons can condense in a state  close to $(B_0^\dag)^N|v\ran$, while free electrons or free holes form a Fermi sea. Still, the Pauli exclusion principle shows up through the norm of the $(B_0^\dag)^N|v\ran$ state  as 
\be
\lan v| B_0^{N} B_0^{\dag N}|v\ran =N!F_N\,.\label{7}
\ee
The $F_N$ factor decreases  when $N$ increases, from $F_1=1$ to $F_N=0$ for $N$ larger than the number of $\vp$ states making the exciton. To understand it, we can say that, due to Pauli blocking, a $B^\dag_0$ exciton added to the one-exciton state $B_0^\dag|v\ran$ finds one fermion pair state occupied by the first exciton. And so on, when more excitons are added. So, excitons can be piled up in the same state, up to a critical amount, above which their fermionic components show up in a dramatic way by canceling the state norm.

The situation is totally different for a dot duo. When the confinement is strong, the carrier states in a dot have a large energy separation, each energy level being at most occupied by two carriers with different spins -- or spin-like indices in the case of holes. As a result, the creation operator of a duo having a well-defined energy is not a sum of fermion pair states as in the case of excitons, but a single fermion pair characterized by the dot level indices
\be
B^\dag_{n_e,n_h}=a^\dag_{n_e}b^\dag_{n_h}\,.
\ee
As a direct consequence, the product of two creation operators for the same duo is equal to zero, $(B^\dag_{n_e,n_h})^2=0$.

This shows that an exciton behaves as a boson when the number of excitons is small compared to the number of fermion pairs that make it, while a duo, which  corresponds to {\it one} fermion pair only, behaves as a fermion from the very first one.

 \textbf{(iii) Coupling to photon}
 
A photon corresponds to a plane wave with momentum $\vQ$ along its propagation direction. It moreover is characterized by a transverse vector associated with its polarization. The photon momentum and polarization are {\it a priori} conserved in the coupled electronic state.

This conservation is easy to achieve for bulk semiconductors. Indeed, the elementary excitations coupled to photons correspond to electron-hole pairs $(\vk_e,\vk_h)$ which, due to Coulomb attraction, can form  bound excitons with center-of-mass momentum $\vK=\vk_e+\vk_h$ and relative-motion index $\nu$. Conservation of the photon momentum imposes the plane-wave photon $\vQ$ to transform into the plane wave for the exciton center of mass, that is, $\vK=\vQ$. Conservation of the photon polarization imposes a constraint on the orbital part of the exciton relative motion. More precisely, in a semiconductor, the upper valence states are characterized by a three-fold index $\lambda=(x,y,z)$. Due to symmetry, a photon with polarization vector along ${\bf x}$ couples a valence electron in a ${x}$  state to a (non-degenerate) state of the conduction band, as this state has opposite parity. Moreover, since photons do not act on spin, the valence electron which goes to the conduction band keeps its $s=\pm1/2$ spin; so, the total spin of the photocreated electron-hole pair reduces to zero. Actually, this simple picture is mixed up by the spin-orbit interaction which provides sizable splittings among the degenerate valence states. As a result, the ($3\times2$) valence states are not  labeled by $(\lambda,s)$ but by quantum indices that can be conveniently taken as $(j,j_z)$ with $j=(3/2,1/2)$ and $-j\leq j_z\leq j$, although $j$ for the valence electrons is fundamentally different from the angular momentum $j$ of an atom, because the potential felt by electrons in a lattice is not spherical but periodic, so that angular momenta have no meaning for them. It is possible to show that the coupling between a circularly polarized photon $\eta=\pm1$ propagating with a momentum $\vQ$ along ${\bf z}$, and a bulk exciton characterized by a center-of-mass momentum $\vQ$, a relative-motion index $\nu$, an electron spin $s=\pm 1/2$ and a hole quantum index $(j,j_z)$, reads as\cite{Combescotbook}
\be
W=\sum_{\vQ,\eta}  \sum_{\nu} \sum_{s,j,j_z}\,\Omega_{\eta,\nu;s,j,j_z}B^\dag_{\vQ,\nu;s,j,j_z}\alpha_{\vQ;\eta}+h.c.\, \label{8}
\ee
where $\alpha^\dag$ is the photon creation operator. The $\Omega_{\eta,\nu;s,j,j_z}$ Rabi coupling  differs from zero for $\eta=s+j_z$ only, due to polarization conservation.

So, a plane-wave photon $\vQ$ fundamentally transforms into a plane-wave bound exciton $\vQ$, and this $\vQ$ photon is restored by the recombination of the $\vQ$ exciton.

When one spatial direction is confined, as in quantum wells, the component of the photon momentum along this direction cannot be conserved: an extended plane wave then transforms into a localized state, which brings a strong reduction of the coupling compared to its bulk value. To mitigate this reduction, quantum wells are commonly put at the field node of a microcavity that enforces the photon momentum to have quantized values comparable with the components of the carrier momenta perpendicular to the well\cite{microcavity}. The same occurs for quantum wires with selected photon momenta comparable with the momenta of the confined carriers\cite{QWR}.

The situation is totally different for QDs because the carrier momenta are confined in three spatial directions. Discrete energy levels for small-size dots indicate strongly confined wave functions. So, the absorption of a photon in a QD  corresponds to transforming the plane-wave photon into a strongly localized electron-hole pair state, thereby making the resulting coupling much smaller than for bulk exciton. The characteristics that are kept from the absorbed photon are its energy that is adjusted to the dot state we want to couple to, and the photon polarization.

\begin{figure}[t]
\begin{center}
\includegraphics[trim=6cm 7.5cm 6cm 5cm,clip,width=2.8in] {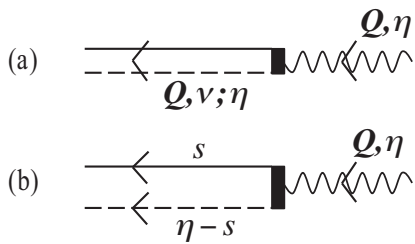}
\end{center}
\vspace{-0.5cm}
\caption{\small Coupling of a photon with momentum $\vQ$ and circular polarization $\eta=\pm1$ (a) to an exciton: the emitted exciton has the same momentum $\vQ$ and polarization $\eta$; (b) to a dot duo made of a $s=\pm1/2$ conduction electron and a hole with index $j_z=\eta-s$: information on the photon momentum $\vQ$ is lost but information on the polarization $\eta$ is kept in the resulting pair.}
\label{fig:1}
\end{figure}

The photon coupling to an exciton and a dot duo are schematized in Fig.~\ref{fig:1}. The momentum $\vQ$ of the photon is lost in the case of a QD.

\textbf{(iv) Photon absorption}

Another major difference between an exciton and a dot duo is the absorption mechanism. Indeed, according to Eq.~(\ref{8}), a photon with momentum $\vQ$ creates an exciton with momentum $\vQ$, the same $\vQ$ photon being re-emitted when the $\vQ$ exciton recombines. This repeated photon-exciton transformation results in a mixed photon-exciton quantum state called polariton, as first noted by Hopfield\cite{Hopfield}. So, the question one must ask concerning  exciton is why a photon is absorbed, that is, not re-emitted anymore?

Actually, the momentum of the exciton coupled to the $\vQ$ photon can change due to collisions with impurities, defects or other excitons. The physical scenario is that a photon $\vQ$ should transform into an exciton $\vQ$. But if the exciton changes its momentum from $\vQ$ to $\vQ'$ on a time scale small compared to the exciton recombination time, the exciton which ultimately recombines has a momentum $\vQ'$ different from $\vQ$ and it emits a $\vQ'$ photon; so, the $\vQ$ photon is not emitted anymore. In other words, it is lost, or absorbed. This scenario  also explains why the photon absorption can be obtained along the Fermi golden rule\cite{Dubin}. Indeed, the Fermi golden rule is known to be valid for transitions toward a continuum of states. In the present problem, a $\vQ$ photon is \textit{a priori} coupled not to a continuum but to a discrete state, namely, the exciton having the same momentum  $\vQ$. The fast collisions suffered by this $\vQ$ exciton induce a kind of continuum made of $\vQ'$ exciton states, which renders the Fermi golden rule effectively valid.

The situation is totally different in the case of a dot duo because the photon momentum $\vQ$ is totally lost in the coupling between a photon and a dot duo (see Fig.~\ref{fig:1}). So, the photon which is re-emitted by this duo has no reason to be the initial photon. As a result, the initial $\vQ$ photon is lost, or absorbed.

All this shows that photon absorption is intrinsic in the case of dot duos, while due to sample imperfections, photon absorption for excitons is restricted to the so-called ``weak coupling" regime.

\textbf{As a conclusion}, we hope to have convinced the reader that electrons and holes in a small quantum dot deserve to be called differently from carriers in a bulk, quantum well and quantum wire, due to severe differences in their intrinsic properties as well as their physics when photons are involved. This is why, instead of exciton, trion and biexciton which  are similar objects in bulk, quantum well and quantum wire, we propose to call them in a dot by the musically charming names, duo, trio and quatuor.

\appendix

\section{Carrier leakage from the dot\label{app:A}} 

Using basic quantum mechanics,  we here show that a conduction electron confined in a small QD has a wave function that spreads more across the potential barriers than a heavy hole. As a consequence, the dot duo appears as ``positively" charged and behaves as an attractive center for another electron. 

We consider a carrier with mass $m$, trapped by a barrier  energy $U$. The lighter the carrier is, the farther it can escape, provided that the barrier energy is not infinite.
To show it, we take a model QD, with dimension $d$ along the growth axis z much smaller than  the other two dot directions. The QD confinement imposes to the carrier a localization energy $\varepsilon$ essentially controlled by $(h-\varepsilon)\varphi(z)=0$, with $h=-(\hbar^{2}/2m)\partial^{2}/\partial z^2 + U(z)$, for $U(z)=-U$ when $-d/2 < z < d/2$ and $U(z) =0$ otherwise. When written in  $\bar{z}=z/d,~ \bar{\varepsilon}=\varepsilon/(\hbar^{2}/2md^2)$ and 
\begin{equation}
\label{A1}
\bar{U} = U \frac{2md^{2}}{\hbar^2}\,,
\end{equation}
the Schr\"{o}dinger equation for a carrier in the dot reduces to 
\begin{equation}
\label{A2}
\Big(-\frac{\partial^{2}}{\partial \bar{z}^{2}}-\bar{U}\, \theta(\bar{z})-\bar{\varepsilon}\Big)\varphi(\bar{z})=0\,,
\end{equation}
with $\theta(\bar{z})=1$ for $-1/2 < \bar{z} < 1/2$. The probability for the carrier to be inside the dot reads as
\begin{equation}
\label{A3}
\mathcal{P}^{(in)}_{\bar U} = \frac{\int_{-1/2}^{1/2}\,\,\mathrm{d}\bar{z}\,\,|\varphi(\bar{z})|^{2}}{\int_{-\infty}^{\infty}\,\mathrm{d}\bar{z}\,\,|\varphi(\bar{z})|^{2}}\,.
\end{equation}
When $\bar{U} \rightarrow \infty$, the carrier is inside the QD and $\mathcal{P}^{(in)}_{\bar U}$ goes to 1, while when $\bar{U} \rightarrow 0$, there is no barrier, and the probability to be inside the dot reduces to the dot volume divided by the (infinite) sample volume, which effectively is zero. 

\begin{figure}[t]
\begin{center}
\includegraphics[trim=0cm 0.5cm 1cm 0cm,clip,width=3.2in]{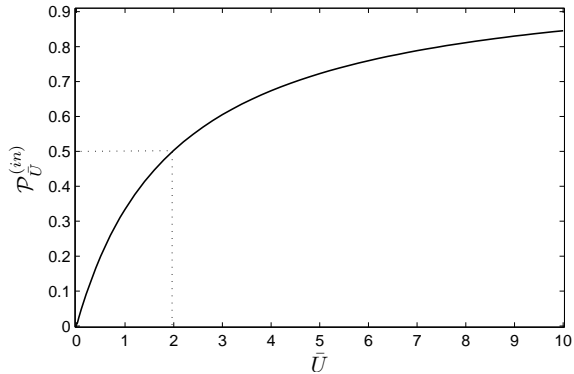}
\caption{Probability for the carrier with mass $m$ to be trapped in a flat dot of width $d$ and  potential barrier $U$, as a function of dimensionless parameter $\bar{U} =U/(\hbar^2/2md^2)$. The probability quickly reaches $1/2$ for $\bar{U}\simeq2$. }
\label{fig:2}
\end{center}
\end{figure}

The curve of Fig.~\ref{fig:2} shows that the probability to be inside the dot increases with $\bar{U}$, that is with $m U$. As a result, a heavy carrier mass can compensate a low barrier. In the case of InAs, the heavy hole mass is $\sim 0.343 m_0$, which is one order of magnitude heavier than the electron mass $\sim 0.027 m_0$ \cite{Kim}, where $m_0$ is the free electron mass. When embedded in GaAs barriers, the gap difference $\Delta E_{gap}$ between the two semiconductors is about $1$ eV. It is commonly accepted that this gap difference is divided into 2/3 for the conduction band and 1/3 for the valence band. So $\bar{U}_{e}/\bar{U}_{hh}\simeq 2 m_{e}/m_{hh} < 1 $. So, due to its small mass, the conduction electron leaks out of the QD more than the heavy hole. This makes the dot occupied by a duo appear as positively charged. So, an electron gains energy by joining the dot with a duo, while a hole loses energy by doing so.


\begin{thebibliography}{99}

\bibitem{Bastard} There are many textbooks on the electronic and optical properties of semiconductor nanostructures. For instance see: G. Bastard, \textit{Wave mechanics applied to semiconductor heterostructures}, Les Editions de Physique, CNRS 1988; J. Davies, \textit{The Physics of low dimensional semiconductors, an introduction}, Cambridge University Press, 1998.
\bibitem{spin}I. \v{Z}uti\'{c}, J. Fabian, and S. Das Sarma, Rev. Mod. Phys. \textbf{76}, 323 (2004).
\bibitem{Michler} A recent overview can be found in \textit{Quantum Dots for Quantum Information Technologies}, ed. Peter Michler, Springer, 2017.
\bibitem{2D} See for instance: S. I. Pokutnyi, Phys. Stat. Sol. \textbf{173}, 607 (1992); S. M. Reimann and M. Manninen,
Rev. Mod. Phys. \textbf{74}, 1283 (2002); G. Wang, A. Chernikov, M. M. Glazov, T. F. Heinz, X. Marie, T. Amand, and B. Urbaszek, Rev. Mod. Phys. \textbf{90}, 021001 (2018); J.-C. Blancon, A. V. Stier, H. Tsai, W. Nie, C. C. Stoumpos, B. Traor\'{e}, L. Pedesseau, M. Kepenekian, F. Katsutani, G. T. Noe, J. Kono, S. Tretiak, S. A. Crooker, C. Katan, M. G. Kanatzidis, J. J. Crochet, J. Even, and A. D. Mohite, Nat. Commun. \textbf{9}, 2254 (2018).
\bibitem{exciton} R. S. Knox, \textit{Theory of Excitons,} Academic Press, New York (1963).
\bibitem{Guillet} M. Combescsot and T. Guillet, Eur. Phys. J. B \textbf{34}, 9 (2003).
\bibitem{Zunger} A. J. Williamson, L. W. Wang, and A. Zunger, Phys. Rev. B \textbf{62}, 12963 (2000); G. Bester, S. Nair, and A. Zunger, Phys. Rev. B \textbf{67}, 161306(R) (2003).
\bibitem{MAD} M. A. Dupertuis, K. F. Karlsson, D. Y. Oberli, E. Pelucchi, A. Rudra, P. O. Holtz, and E. Kapon,
Phys. Rev. Lett. \textbf{107}, 127403 (2011).
\bibitem{Efros} A. L. Efros and A. L. Efros, Fiz. Tekh. Poluprovodn. \textbf{16}, 1209 (1982); Sov. Phys. Semicond. \textbf{16}, 772 (1982).
\bibitem{Hanamura} E. Hanamura, Phys. Rev. B \textbf{37}, 1273 (1988).
\bibitem{Takagahara} T. Takagahara, Phys. Rev. B \textbf{36}, 9293 (1987).
\bibitem{Que} W. Que, Phys. Rev. B \textbf{45}, 11036 (1992).
\bibitem{Andreani} L. C. Andreani, G. Panzarini, and J.-M. G\'{e}rard, Phys. Rev. B \textbf{60}, 13276 (1999).
\bibitem{Stobbe} S. Stobbe, P. T. Kristensen, J. E. Mortensen, J. M. Hvam, J. M\o rk, and P. Lodahl, Phys. Rev. B \textbf{86}, 085304 (2012).
\bibitem{Shiau} S. -Y. Shiau, M. Combescot, and Y.-C. Chang, Phys. Rev. B \textbf{86}, 115210 (2012).
\bibitem{trion} A.J. Shields, J.L. Osborne, M.Y. Simmons, M. Pepper, and D.A. Ritchie, Phys. Rev. B \textbf{52}, R5523(R) (1995); S. Glasber, S. Finkelstein, H. G. Shtrikman, and I. Bar-Joseph, Phys. Rev. B \textbf{59}, R10425 (1999).
\bibitem{biexciton} R. C. Miller, D. A. Kleinman, A. C. Gossard, and O. Munteanu, Phys. Rev. B \textbf{25}, 6545 (1982).
\bibitem{BX} S. Rodt, R. Heitz, A. Schliwa, R. L. Sellin, F. Guffarth, and D. Bimberg, Phys. Rev. B \textbf{68}, 035331 (2003).
\bibitem{X-X+} G. Bester, and A. Zunger, Phys. Rev. B \textbf{68}, 073309 (2003); A. Schliwa, M. Winkelnkemper, and D. Bimberg, Phys. Rev. B \textbf{79}, 075443 (2009); M, Zieli\'{n}ski, K. Go?asa, M. R. Molas, M. Goryca, T. Kazimierczuk, T. Smole\'{n}ski, A. Golnik, P. Kossacki, A. A. L. Nicolet, M. Potemski, Z. R. Wasilewski, and A. Babi\'{n}ski, Phys. Rev. B \textbf{91}, 085303 (2015).
\bibitem{Finley} R. Warburton \textit{et al}, Nature (London) \textbf{405}, 926 (2000); J. J. Finley, P. W. Fry, A. D. Ashmore, A. Lema\^{i}tre, A. I. Tartakovskii, R. Oulton, D. J. Mowbray, M. S. Skolnick, M. Hopkinson, P. D. Buckle, and P. A. Maksym, Phys. Rev. B \textbf{63}, 161305(R) (2001).
\bibitem{Regelman} D. V. Regelman, E. Dekel, D. Gershoni, E. Ehrenfreund, A. J. Williamson, J. Shumway, A. Zunger, W. V. Schoenfeld, and P. M. Petroff, Phys. Rev B \textbf{64}, 165301 (2001). 
\bibitem{Combescotbook} M. Combescot and S.-Y. Shiau, \textit{Excitons and Cooper Pairs}, Oxford University Press (Oxford, 2016).
\bibitem{microcavity} R. Houdr\'{e}, R. P. Stanley, U. Oesterle, M. Ilegems, and C. Weisbuch, Phys. Rev. B \textbf{49}, 16761 (1994).
\bibitem{QWR} E. Wertz, L. Ferrier, D. D. Solnyshkov, R. Johne, D. Sanvitto, A. Lema\^{i}tre, I. Sagnes, R. Grousson, A. V. Kavokin, P. Senellart, G. Malpuech, and J. Bloch, Nat. Phys. \textbf{6}, 860 (2010); P. Lodahl, S. Mahmoodian, and S. Stobbe, Rev. Mod. Phys. \textbf{87}, 347 (2015).
\bibitem{Hopfield} J. J. Hopfield, Phys. Rev. \textbf{112}, 1555 (1958).
\bibitem{Dubin} F. Dubin, M. Combescot and B. Roulet, Eur. Phys. Lett. \textbf{69}, 931 (2005).
\bibitem{Kim} Y. Kim, K. Hummer, and G. Kresse, Phys. Rev. B \textbf{80}, 035203 (2009).



\end{thebibliography}
\end{document}